\documentclass[%
preprint,
 amsmath,amssymb,
 aps,
]{revtex4-1}

\usepackage{graphicx}
\usepackage{dcolumn}
\usepackage{bm}
\usepackage{comment}
\usepackage{float}
\usepackage{amsmath}
\usepackage{cases}
\usepackage{listings}
\usepackage{xcolor}
\usepackage{subfigure}
\usepackage[colorlinks]{hyperref}
\usepackage{footmisc}
\begin{document}
\renewcommand{\thefootnote}{\fnsymbol{footnote}}

\preprint{APS/123-QED}

\title{Observing Nonclassical Paths by Measuring Bohmian velocities in Multi-Slit Interferometer with Electron}

\author{Xiuzhe Luo\footnote[1]{These authors contributed equally to this work.}}
\author{Zhiyuan Wei\footnotemark[1]}
\author{Yong-Jian Han}
\author{Chuan-Feng Li}
 \email{cfli@ustc.edu.cn}
\author{Guang-Can Guo}
 \affiliation{Key Laboratory of Quantum Information, University of Science and Technology of China, CAS, Hefei, Anhui 230026, China.}
\affiliation{Synergetic Innovation Center of Quantum Information \& Quantum Physics, University of Science and Technology of China, Hefei, Anhui 230026, China.}

\date{\today}

\begin{abstract}
Superposition principle of wave function in multi-slit interference experiment has been widely accepted by many quantum mechanics textbooks, however the expression ${\psi _{AB}} = {\psi _A} + {\psi _B}$ is not strictly hold due to multipath interference. The nonclassical paths in Feynman path integral formalism provide a quantitatively way to analyze the this effect. In this work, we apply Bohmian mechanics to the measurement of multipath interference by analyzing the contribution of nonclassical paths to the Bohmian velocities in electron's multi-slit interferometer. The results show that the nonclassical paths lead to a relative deviation of order $10^{-3}$ in electron's Bohmian velocities, which is observable by weak measurement approach.
\end{abstract}

\maketitle


\section{Introduction}

The superposition principle of wave function is one of the most essential feature of quantum mechanics. From this principle, physicists are able to understand many phenomena like the interference pattern of multi-slit interference experiment. Basically, for triple-slit interference experiment with slits $A$, $B$, $C$, it is usually assumed that we have 
\begin{equation} \label{eq:1}
	\phi_{ABC} = \phi_A + \phi_B + \phi_C
\end{equation}

on observational screen\cite{cohen1991quantum,shankar2012principles}.

Though this assumption is widely reached in various quantum mechanics textbooks, actually this expression is not a precise description. The cases of opening different combinations of slit $A$, $B$ and $C$ actually correspond to different boundary conditions \cite{yabuki1986feynman,de2012analysis}.

In 1994, R.Sorkin proved that if the \eqref{eq:1} holds, the multipath interference term in triple-slit interference experiment should be zero everywhere on the observing plane \cite{sorkin1994quantum}. This result provide us a way to quantitative way to see the deviation to \eqref{eq:1}. In 2010, U.Sinha et al. bounded the magnitude of the multipath interference to less than $10^{-2}$ of the regular expected interference, at several detector positions \cite{sinha2010ruling}.

A numerical simulation\cite{de2012analysis} on analyzing the classical case of multipath interference in three-slits experiments have shown the deviation from expression \eqref{eq:1}. In 2014, R.Sawant et al. used the nonclassical paths, which are paths in Feynman path integral that don’t extremize the classical action, to estimate the magnitude of multipath interference in three-slits experiment \cite{sawant2014nonclassical} . A famous demonstration of nonclassical paths is related to the Aharonov-Bohm effect\cite{aharonov1959significance}.

\begin{figure}[h!]
	\centering
	\includegraphics[width=0.7\textwidth]{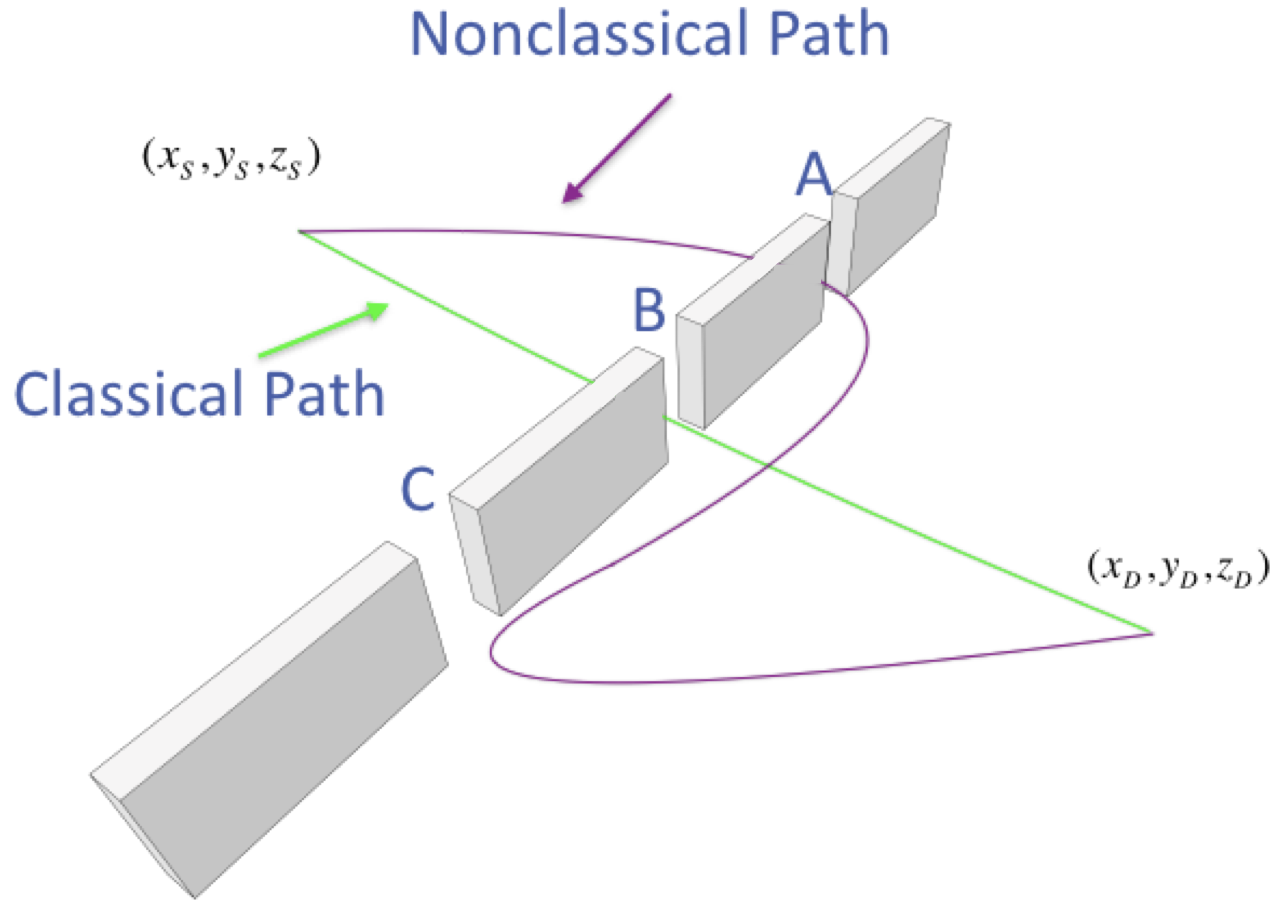}
	\caption{(color online). Schematic Graph of path integrals in 3-slits interference experiment. The green line is a classical path, the purple line is a nonclassical path starts and ends at the same point with the classical path. This nonclassical path corresponds to part of the 2-order approximation term in the expansion of Feynman propagator $K$.}
	\label{skem1}
\end{figure}

  The experiment approach\cite{sinha2010ruling} is a feasible way to measure the nonclassical paths effect by observing the interference fringes. However it requires multi-setting measurement by choosing different combinations of opening slits, which decreases the stability and produces additional measurement error. In this work, we propose a new way to observe the nonclassical paths effect in electron's multi-slit interferometer by measuring Bohmian velocities.
  
  Bohmian mechanics is an interpretation of quantum theory \cite{bohm1952p1,bohm1952p2}. In Bohmian mechanics, there is a “hidden” variable that exists even when unobserved. In Bohmian mechanics, the evolution of a system’s configurations over time is defined by the wave function via a guiding equation, and thus this theory is called Pilot-Wave Theory. Classical pilot wave phenomena have been found since 2005 in a floating droplet on vertical vibrating bath system by Y.Couder et al.\cite{couder2005dynamical}, later many interesting phenomena like single particle diffraction and level splitting also found in this system\cite{couder2006single,eddi2009unpredictable,eddi2012level,harris2013wavelike}. In 2013, Aephraim M. Steinberg et al. measured the average trajectories of a single photon in two-slit interferometer using weak measurement, and the result reproduce trajectories predicted by Bohmian mechanics \cite{kocsis2011observing}.

Since the nonclassical paths contribute to the probability amplitude in path integral formalism, their contribution will also affect the trajectories predicted by Bohmian mechanics, thus if we measure the average trajectories of a particle in a three-slit interferometer, such a nonclassical paths effect should be expected. 

In the following passage, we will analytically and numerically calculate the nonclassical paths effect on Bohmian velocities and trajectories in electron's three-slits interferometer.

\section{\label{sec:level1}Analytical Calculation}
In this section, we do analytical analysis for the wave function and velocity in electron's multi-slit interference experiment under two different conditions: (1) not counting nonclassical paths effect; (2) counting nonclassical paths effect. The nonclassical paths in our analysis contain the paths which enter a slit and go to another slit, then directly go to the observing plane\cite{sinha2015superposition}.
\subsection{Contribution of classical paths to the probability amplitude}

We use the Feynman path integral in two dimensions to calculate the probability amplitude of path integrals for triple slits,

\begin{figure}[h!]
	\centering
	\includegraphics[width=0.8\textwidth]{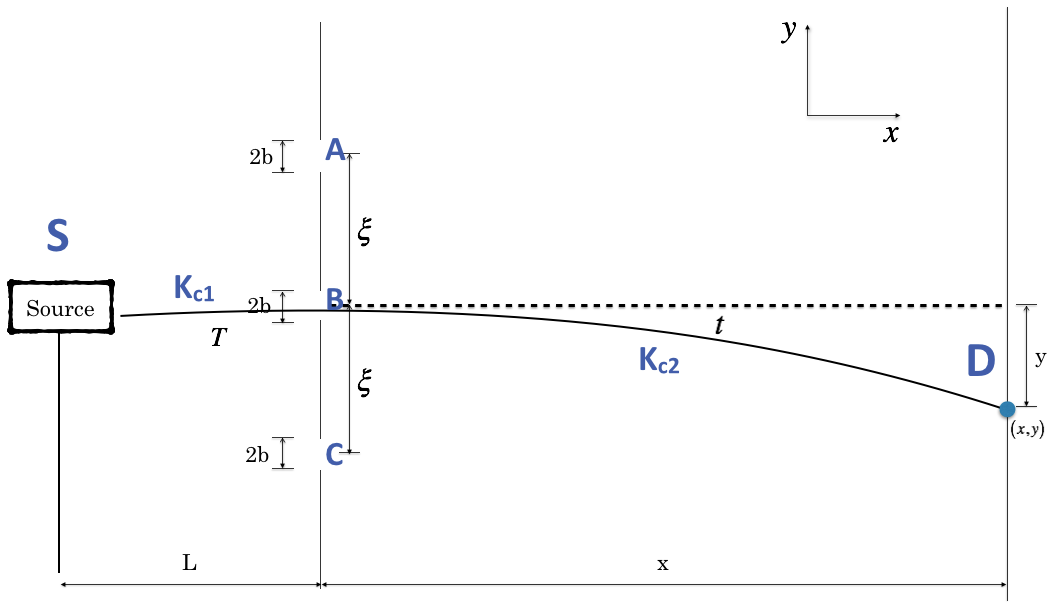}
	\caption{(color online). Representation of symbols in classical path calculation.}
	\label{skem_cl}
\end{figure}

Fig.\ref{skem_cl} shows a classical path in triple-slit interference system which starts from the source $S$ and ends at the observing plane $D$. $L$ is the distance from source to the grating plane, while $x$ is the distance from the grating plane to the observing plane. In the free propagation case, the electron's velocity in $x$ direction is approximately a constant $V_x$, so $T = L/V_x$ is the transport time from source to the grating plane, $t = x/V_x$ is the transport time from the grating plane to the observing screen. $\xi$ is the distance from the given slit to the center of the grating plane (the dashed line). Electron's mass is denoted by $m$ and the width of the slits are denoted by $2b$.

Here we show how to calculate the propagator of a classical path that cross slit $B$. For free propagation, the propagator from source to a given slit can be written in following form \cite{feynman2005quantum}:

\begin{equation} \label{eq:2}
K_{c1}(S,\xi_B) =(\frac{2\pi i \hbar T}{m})^{-\frac{1}{2}}\exp{\frac{im}{2\hbar}[\frac{L^2+(\xi_B+\delta y)^2}{T}]}	
\end{equation}

\begin{equation} \label{eq:3}
K_{c2}(\xi_B,D) =(\frac{2\pi i \hbar t}{m})^{-\frac{1}{2}}\exp{\frac{im}{2\hbar}[\frac{x^2+(y-\xi_B-\delta y)^2}{t}]}	
\end{equation}

Where the relevant $y$-direction distance to center of the slit is denoted by $\delta y$, and it will be eliminated by integration over the $y$-direction of the slit.

$K_{c1}$ and $K_{c2}$ correspond to two parts of this path(see in Fig.\ref{skem_cl}). So the contribution of this classical path to the wave function is:

\begin{equation} \label{eq:4}
\psi_c (x,y,{\xi _B}) = \int_{ - b}^b {d(\delta y)G\left( {\delta y} \right){K_{c1}}} (S,{\xi _B}){K_{c2}}({\xi _B},D)
\end{equation}

The slit transmission probability is denoted by $G\left( {\delta y} \right)$. Here we apply Gaussian slits\cite{feynman2005quantum} in our calculation:
\begin{equation} \label{eq:5}
	G(\delta y) = exp( - {(\delta y)^2}/2b)
\end{equation}

where $b$ is the width of the slit. The propagator becomes
\begin{equation} \label{eq:6}
\psi_c (x,y,{\xi _B}) = \int_{ - \infty }^\infty  {d(\delta y)\frac{m}{{2\pi i\hbar }}} \exp \left( { - \frac{{{{(\delta y)}^2}}}{{2b}} + \frac{{im}}{{2\hbar }}[\frac{{({L^2} + {{({\xi _B} + \delta y)}^2})}}{T} + \frac{{{x^2} + {{(y - {\xi _B} - \delta y)}^2}}}{t}]} \right)
\end{equation}

The result of this integral is:
\begin{equation} \label{eq:7}
\psi_c (x,y,\xi_B ) = \sqrt {\frac{m}{{ - 2\pi i\mu \hbar Tt}}} \exp (\frac{{im}}{{2\hbar }}[\frac{{{L^2} + {\xi_B ^2}}}{T} + \frac{{{x^2} + {{(y - \xi_B )}^2}}}{t} - \frac{{{{(\frac{\xi_B }{T} + \frac{{\xi_B  - y}}{t})}^2}}}{\mu }])	
\end{equation}

where \(\mu = \frac{1}{T}+\frac{1}{t}-\frac{\hbar}{imb^2}\).

So we have the wave function of multiple slits:
\begin{equation} \label{eq:8}
\Psi_c (x,y) = \sum\limits_i {\psi_c (x,y,{\xi _i})}	  \qquad   (i=A,B,C,...)
\end{equation}

\newpage
\subsection{Contribution of Nonclassical paths to the probability amplitude}

\begin{figure}[h!]
	\centering
	\includegraphics[width=0.7\textwidth]{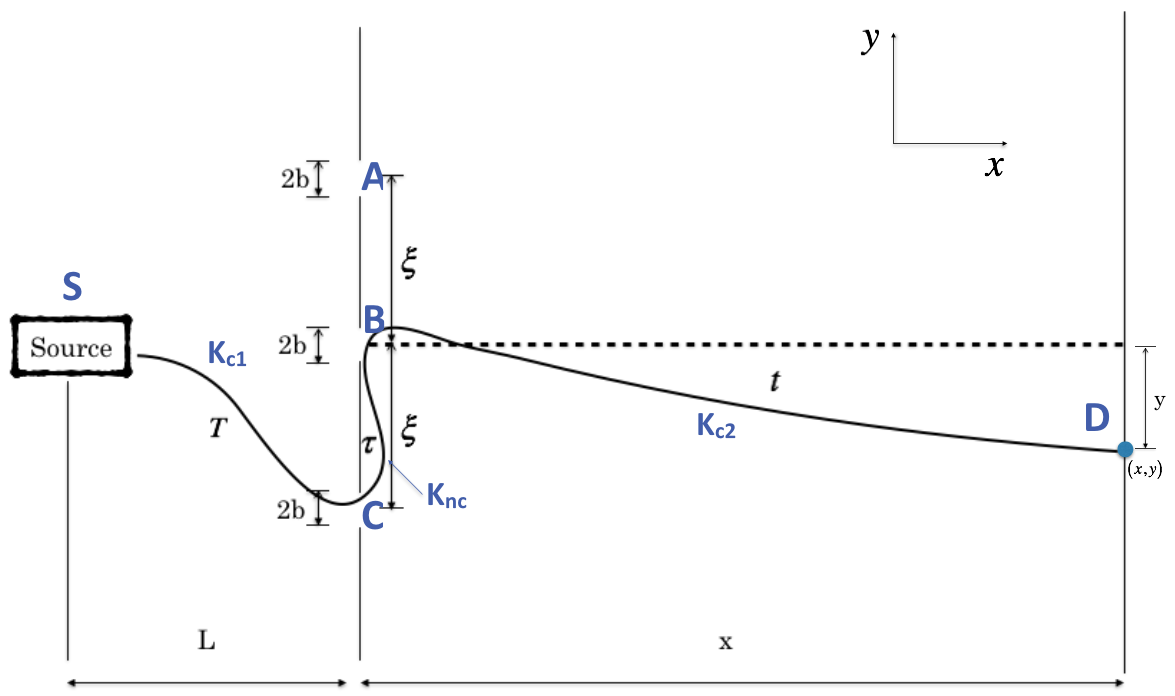}
	\caption{(color online). Representation of symbols in nonclassical path calcuation.}
	\label{skem2}
\end{figure}

To analyse the effect of nonclassical path in the Bohmian velocity, we analytically calculate the propagator of following path (see in Fig.\ref{skem2}):

 Define $\tau$ as the transport time from slit $C$ to slit $B$ , $\xi_C$ as the $y$-position of slit $C$ , $\xi_B$ as the $y$-position of slit $B$. The propagator between two slits $C,B$ can be written as : 

\begin{equation} \label{eq:9}
K_{nc}(C,B) = (\frac{2\pi\hbar T}{m})^{-\frac{1}{2}}\exp{\frac{im}{2\hbar}[\frac{(\xi_C+\delta_{yc}-\xi_B-\delta_{yb})^2}{\tau}]}	
\end{equation}

We can estimate the transport time $\tau$ using the Heisenberg's uncertainty relation (similar to \cite{da2016gouy}):
\begin{equation} \label{eq:10}
	v_y = \frac{\hbar}{(\xi_C-\xi_B)\cdot\sqrt{2}m}
\qquad \quad
\tau = \frac{(\xi_C-\xi_B)^2\cdot\sqrt{2}m}{\hbar}
\end{equation}

For the nonclassical path in Fig.\ref{skem2}, its contribution to the wave function ${\psi _{nc}}$ can be calculated by integrating over the slit $C$ and $B$ in $y$ direction:
\begin{equation} \label{eq:11}
{\psi _{nc}}(x,y,{\xi _i},{\xi _j}) = \int_{ - b}^b d {\delta _{yc}}\int_{ - b}^b d {\delta _{yb}}{K_{c1}}(S,C){K_{nc}}(C,B){K_{c2}}(B,D)
\end{equation}

After applying the Gaussian slits transmission rate, we can obtain the result of this integral:
\begin{equation} \label{eq:12} 
	\begin{array}{*{20}{l}}
{{\psi _{nc}}(x,y,{\xi _i},{\xi _j}) = \frac{{{{(\frac{m}{{i\hbar }})}^{3/2}}}}{{2\sqrt {2\pi tT\tau \eta \beta } }} \cdot \exp [\frac{{im{{({\xi _j} - y)}^2}}}{{2\hbar t}} + \frac{{i{L^2}m}}{{2\hbar T}} + \frac{{im\xi _i^2}}{{2\hbar T}} + \frac{{im{x^2}}}{{2\hbar t}} - \frac{{{m^2}\xi _i^2}}{{4{\hbar ^2}{T^2}\eta }}}\\
+ {\frac{{{{(\frac{{im{\xi _j}}}{{\hbar t}} - \frac{{imy}}{{\hbar t}} + \frac{{{m^2}({\xi _j} - {\xi _i})}}{{2{\hbar ^2}{\tau ^2}\eta }} + \frac{{im({\xi _j} - {\xi _i})}}{{\hbar \tau }} - \frac{{{m^2}{\xi _i}}}{{2{\hbar ^2}\eta \tau T}})}^2}}}{{4\beta }} - \frac{{{m^2}{{({\xi _i} - {\xi _j})}^2}}}{{4{\hbar ^2}{\tau ^2}\eta }} + \frac{{im{{({\xi _i} - {\xi _j})}^2}}}{{2\hbar \tau }} - \frac{{{m^2}(\xi _i^2 - {\xi _i}{\xi _j})}}{{2{\hbar ^2}\eta T\tau }}]}
\end{array}
\end{equation}

where,
\begin{equation} \label{eq:13}
	\begin{aligned}
&\eta = \frac{1}{2b^2}-\frac{im}{2\hbar T}-\frac{im}{2\hbar\tau}\\
&\beta = \frac{1}{2b^2}-\frac{im}{2\hbar t}-\frac{m^2}{4\hbar^2(-\frac{1}{2b^2}+\frac{im}{2\hbar T}+\frac{im}{2\hbar\tau})\tau^2}-\frac{im}{2\hbar\tau}    
\end{aligned}
\end{equation}

So the contribution of nonclassical paths to the wave function is:
\begin{equation} \label{eq:14}
\begin{array}{l}
{\Psi _{nc}}(x,y) = \sum\limits_{ij} {{\psi _{nc}}\left( {x,y,{\xi _i},{\xi _j}} \right)} \qquad (i \ne j)
\end{array}
\end{equation}

\subsection{The change on Bohmian velocity caused by nonclassical paths effect}

We can separately calculate the Bohmian velocity of $y$ direction using guiding equation \eqref{eq:15} with wave function $\psi_c$ and $\psi_{nc}$ to see the nonclassical paths effect.

The Bohmian velocity with only classical paths:

\begin{equation} \label{eq:15}
{v_c} = \frac{\hbar }{m}Im(\frac{{\Psi _c^*\nabla {\Psi _c}}}{{|{\Psi _c}{|^2}}})
\end{equation}

Now separate the wave function to classical and nonclassical parts, and obtain the  Bohmian velocity $v_{nc}$ with both classical and nonclassical paths:

\begin{equation} \label{eq:16}
\Psi(x,y)=\Psi_c+\Psi_{nc}	
\end{equation}
 
\begin{equation} \label{eq:17}
\begin{aligned}
&v_{nc}=\frac{\hbar}{m}Im(\frac{(\Psi_c+\Psi_{nc})^*\nabla(\Psi_c+\Psi_{nc})}{|\Psi_c+\Psi_{nc}|^2})\\
&=\frac{\hbar}{m}Im(\frac{\Psi_c^*\nabla\Psi_c+\Psi_{nc}^*\nabla\Psi_c+\Psi_c^*\nabla\Psi_{nc}+\Psi_{nc}^*\nabla\Psi_{nc}}{|\Psi_c+\Psi_{nc}|^2})
\end{aligned}
\end{equation}

\section{Numerical Analysis}


With the expression of wave function \eqref{eq:16} and Bohmian velocity \eqref{eq:15},\eqref{eq:17}, we can numerically calculate the contribution of nonclassical paths effect on Bohmian velocity using finite-difference methods.

\begin{figure}[h!]
  \centering
  \includegraphics[scale=0.5]{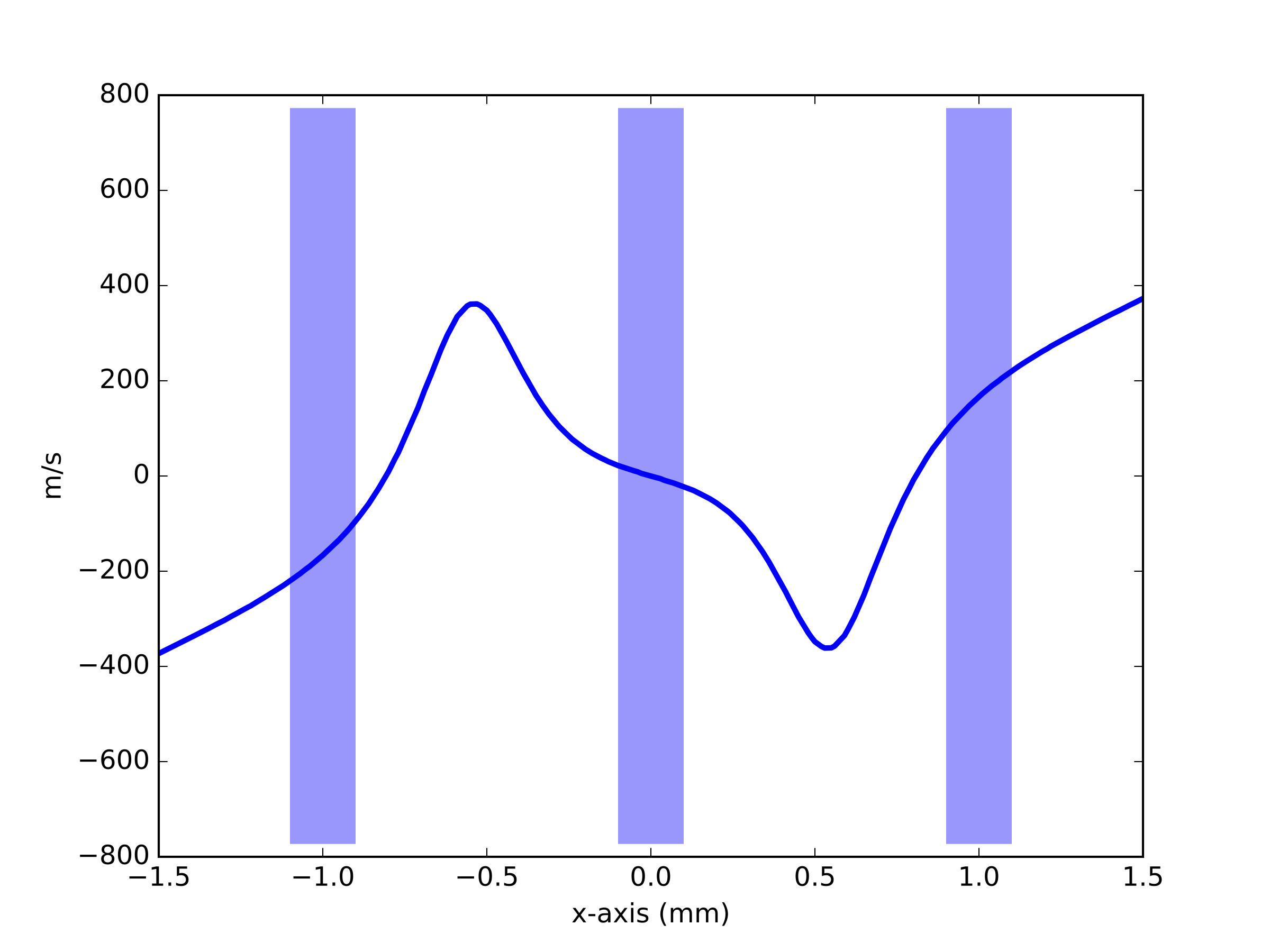}
  \caption{(color online). Electron's $y$-direction Bohmian velocity ${v_yc}$ as a function of detector position $y$ in triple-slit interferometer. The distance between two neighboring slits is $\xi  = {1 \cdot 10^{ - 6}}m$, the width of each slit is $2b = {1 \cdot {10^{ - 7}}}m$. Electron's transport velocity in this simulation is $V_x=1.3\times 10^8$.}
  \label{fig:classical_3}
\end{figure}

\begin{figure}[h!]
  \centering
  \includegraphics[scale=0.5]{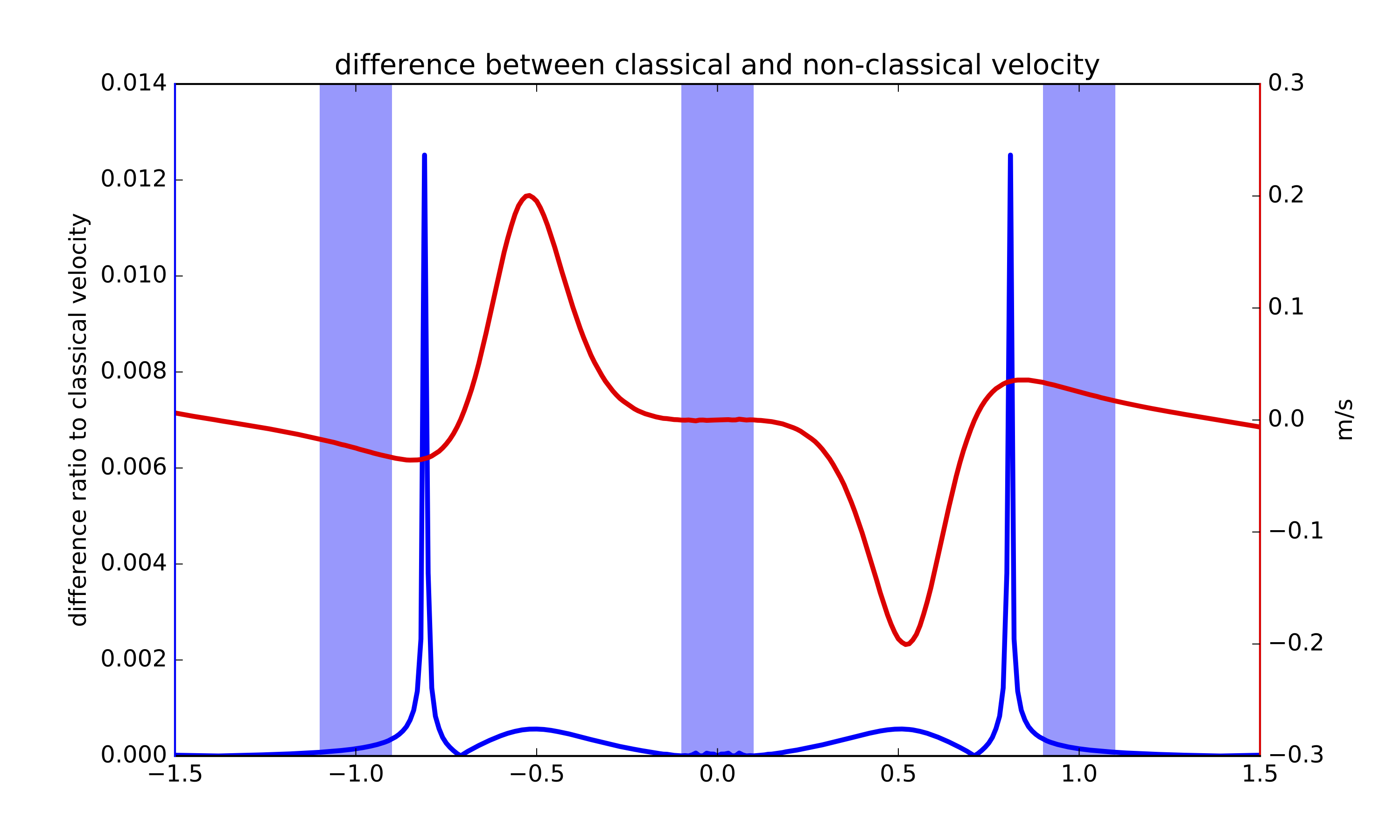}
  \caption{(color online). \textbf{(Red)} The difference of Bohmian velocity ${v_{ync}} - {v_{yc}}$ as a function of detector position $y$ in triple-slit interferometer. \textbf{(Blue)} The relative difference $\frac{{\left| {{v_{ync}} - {v_{yc}}} \right|}}{{\left| {{v_{yc}}} \right|}}$ as a function of detector position $y$ in triple-slit interferometer. The parameters are the same with Fig.\ref{fig:classical_3}.}
  \label{fig:diff_3}
\end{figure}

Fig.\ref{fig:classical_3} shows the Bohmian velocity in $y$-direction ${v_{yc}}$ as a function of detector position $y$. The Bohmian velocity ${v_{nc}}$ which counting the nonclassical paths also looks similar due to the difference is small. The deviation of Bohmian velocity caused by the nonclassical paths effect is shown in Fig.\ref{fig:diff_3}.



The main feature of Fig.\ref{fig:diff_3} are the two sharp spikes around $\left| y \right| = 0.9\mu m$ and finite maximum peaks in around $\left| y \right| = 0.5\mu m$. By comparing with the Fig.\ref{fig:classical_3}, we see that the spikes represent the region where the ${v_{yc}}$ go to zero, and the finite maximum peak represent the peak of the red curve, which make sense for experimental measurement.

\begin{figure}[h!]
  \centering
  \includegraphics[scale=0.5]{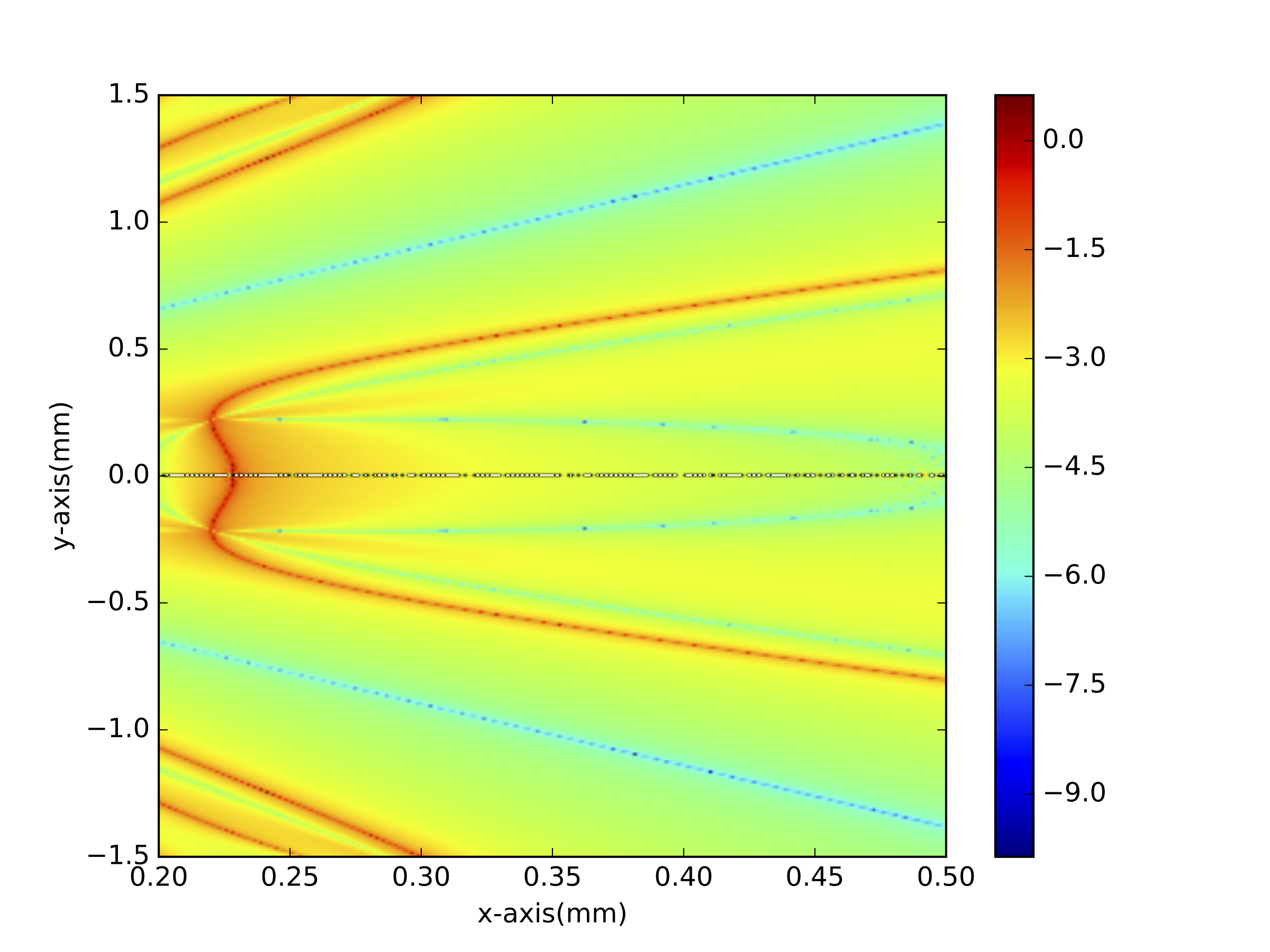}
  \caption{(color online). Density plot of relative difference in $y$-direction Bohmian velocity $\frac{{\left| {{v_{ync}} - {v_{yc}}} \right|}}{{\left| {{v_{yc}}} \right|}}$ for various position $\left( {x,y} \right)$ in triple slits interferometer. The parameters are the same with Fig. \ref{fig:classical_3}.}
  \label{fig:log10}
\end{figure}

The density plot of this relative difference is shown in Fig.\ref{fig:log10}. From this graph, we see the nonclassical paths effect become stronger when near to the grating plane, which is consistent with the result in \cite{sinha2015superposition}. The  maximal relative difference $\frac{{\left| {{v_{ync}} - {v_{yc}}} \right|}}{{\left| {{v_{yc}}} \right|}}$ has the order of ${10^{ - 3}}$.

Previous work \cite{kocsis2011observing} already shows that we can measure the momentum in trajectory of photon using weak measurement, and the result is consistent with the prediction of Bohmian mechanics. Thus the Bohmian velocity here can be measured by similar experiment to see the nonclassical paths effect.

\section{Conclusion}

In conclusion, we analytically and numerically calculated the contribution of nonclassical paths effect in electron's Bohmian velocity in multi-slit interference, showing that the nonclassical paths effect cause a deviation on the Bohmian velocity with a maximum order of ${10^{ - 3}}$. Our work shows the application of the Bohmian mechanics on the measurement of multi-order interference, and we suggest that similar weak measurement experiment can be done to quantify this nonclassical paths effect.

\section{Acknowledgement}
This work was supported by the National Natural Science Foundation of China (Nos. 61327901, 11274289, 11325419), the Strategic Priority Research Program (B) of the CAS (Grant No. XDB01030300), Key Research Program of Frontier Sciences, CAS (No. QYZDY-SSW-SLH003).

\bibliography{reference}

\end{document}